# The cost of a future low-carbon electricity system without nuclear power for Sweden

Xiaoming Kan[a, *]   Fredrik Hedenus[a]   Lina Reichenberg[a, b]

a Department of Space, Earth and Environment, Chalmers University of Technology, Gothenburg, Sweden

b Department of Mathematics and Systems Analysis, Aalto University, Helsinki, Finland

**Abstract**

To achieve the goal of deep decarbonization of the electricity system, more and more variable renewable energy (VRE) is being adopted. However, there is no consensus among researchers on whether the goal can be accomplished without large cost escalation if nuclear power is excluded in the future electricity system. In Sweden, where nuclear power generated 41% of the annual electricity supply in 2014, the official goal is 100% renewable electricity production by 2040. Therefore, we investigate the cost of a future low-carbon electricity system without nuclear power for Sweden. We model the European electricity system with a focus on Sweden and run a techno-economic cost optimization model for capacity investment and dispatch of generation, transmission, storage and demand-response, under a $CO_2$ emission constraint of 10 g/kWh. Our results show that there are no, or only minor, cost benefits to reinvest in nuclear power plants in Sweden once the old ones are decommissioned. This holds for a large range of assumptions on technology costs and possibilities for investment in additional transmission capacity.  We contrast our results with the recent study that claims severe cost penalties for not allowing nuclear power in Sweden and discuss the implications of methodology choice.

## 1. Introduction

The European Commission has presented its strategic long-term vision for a climate-neutral economy by 2050 [1] and has set the target of reducing greenhouse gas emissions to 80-95% below 1990 levels by 2050 [2]. According to the EU Roadmap 2050 [3], the power sector is expected to mitigate nearly all its $CO_2$ emissions by 2050 and meanwhile contribute to carbon reduction in the transport and heating sectors. To achieve this goal, more and more wind- and solar power are invested in Europe for electricity supply [4-6]. In the case of Sweden, the government has set a goal of 100% renewable power production for the electricity sector by 2040 [7]. Currently, nuclear power accounts for 41% of the annual electricity production [8], however the nuclear fleet is aging and decommissioning is planned in the coming decades for economic reasons. This has spurred a political discussion about replacing the old nuclear reactors with new ones. Nuclear power is facing an uncertain future in the transition towards a low-carbon electricity system in Europe due to the risk of radiation leakage, social acceptance, and high investment cost, among other factors. Germany, Belgium, and Switzerland have decided to phase out nuclear power, while Finland and France are building new nuclear power plants.

---

[*] Corresponding author
 Email address: kanx@chalmers.se (Xiaoming Kan)

The cost difference for decarbonizing the electricity system with and without nuclear power has been subject to recent debate in the scientific community [9-12]. Some studies show that excluding nuclear power increases the electricity system cost modestly [13-15], while others claim that the increase in cost is substantial [16, 17]. Jägemann et al. [13] investigated the decarbonization pathways of the European electricity sector and found that the total electricity system cost, together with the cost of decarbonization, would increase by 11% if nuclear power and carbon capture and storage (CCS) were excluded. Zappa et al. [14] evaluated the cost of a 100% renewable power system for Europe and found that the system cost would be 30% higher than a carbon-neutral electricity system which excludes nuclear and CCS. Pattupara and Kannan [15] analyzed the low-carbon electricity pathways in Switzerland and its neighboring countries and showed that the net electricity system cost in the deep decarbonization scenario would increase by 15% if nuclear power were not included. In stark contrast, Buongiorno et al. [16] found that excluding nuclear power would double or triple the average electricity cost for deep decarbonization. Similarly, Sepulveda et al. [17] concluded that firm low-carbon resources, such as nuclear power, might reduce the electricity cost by 10%–62% across fully decarbonized cases. Many other studies have analyzed a renewable future electricity system (without nuclear) in Europe and found that the low-carbon electricity system can be achieved with modest cost increase as compared with the current cost [5, 6, 18-20].

In the case of Sweden, several recent publications [21-23] have evaluated the economic impact of nuclear power exiting the Swedish electricity system and investigated the potential options to replace nuclear power. Hong et al. [21] assessed the cost of phasing out nuclear power in Sweden. Their key finding was that if wind and solar were to replace nuclear power, the average electricity cost[1] would be 303 $/MWh, i.e., around five times higher than the current electricity cost. In comparison, the International Energy Agency (IEA) [22] analyzed the carbon-neutral scenario for the Nordic region and found a cheaper (67 $/MWh for electricity price) electricity system without nuclear for Sweden. Söder

---

[1] Their study did not consider transmission cost, so the cost is based only on investment- and running costs.

[23] investigated the required load balancing and transmission capacity expansion for 100% renewable power production in Sweden and showed that the nuclear power could be replaced by 16 GW wind power and 9 GW solar power, considerably less than that (154 GW wind power) estimated by [21].

Concerning the large difference in the results of studies [21-23] on Sweden, it is important to understand the impact on system cost of not including nuclear power in the future Swedish low-carbon electricity system. We first note two methodological weaknesses in some previous studies [21, 23], including 1) Lack of representation of electricity trade and 2) Lack of system optimization. Söder [23] modeled Sweden as isolated from neighboring countries without international trade, while Hong et al. [21] kept the annual trade at the present level. It is well established in the literature that interregional trade is the key variation management strategy for wind power [4-6, 24]. Therefore, the availability of interconnecting transmission grids has a large effect on system cost. With respect to system optimization, Hong et al. [21], used a heuristic approach to determine the wind and solar capacities to substitute nuclear power without optimizing the whole system. Similarly, Söder [23] did not economically optimize the investment and dispatch of the technology palette for Sweden.

To make a comprehensive analysis of the cost for the electricity system without nuclear power in Sweden and tackle the limitations listed above, we expand the system boundary by including the possibility of trading and investing in transmission capacity. We further develop a techno-economic cost optimization model with high temporal and spatial resolution for the Swedish and European electricity system to address the following questions:

1) What is the cost of a future low-carbon electricity system without nuclear power for Sweden, given the present interconnecting transmission capacities within Sweden and to neighboring countries?
2) How is the cost affected if additional investment in transmission within Sweden and to other countries is allowed?

The paper is organized as follows: The model and input data are introduced in Section 2. In Section 3, the modelling results are presented in terms of net system cost, capacity invested, and electricity generated. The results are then discussed and compared to other studies in the literature in Section 4 and Section 5 concludes. The model-specific code, input data, and output data will be available online to further enhance the transparency and reproducibility of the results.

## 2. Methods

In this study, a future interconnected European electricity system is modeled for the year 2045 with hourly time resolution given a cap on $CO_2$ emission expressed in g $CO_2$ per kWh of electricity demand.

The economic performance of the Swedish electricity system with and without nuclear is evaluated with the nodal net average system cost, which we introduce below. An overview of the method is presented in Fig. 1.

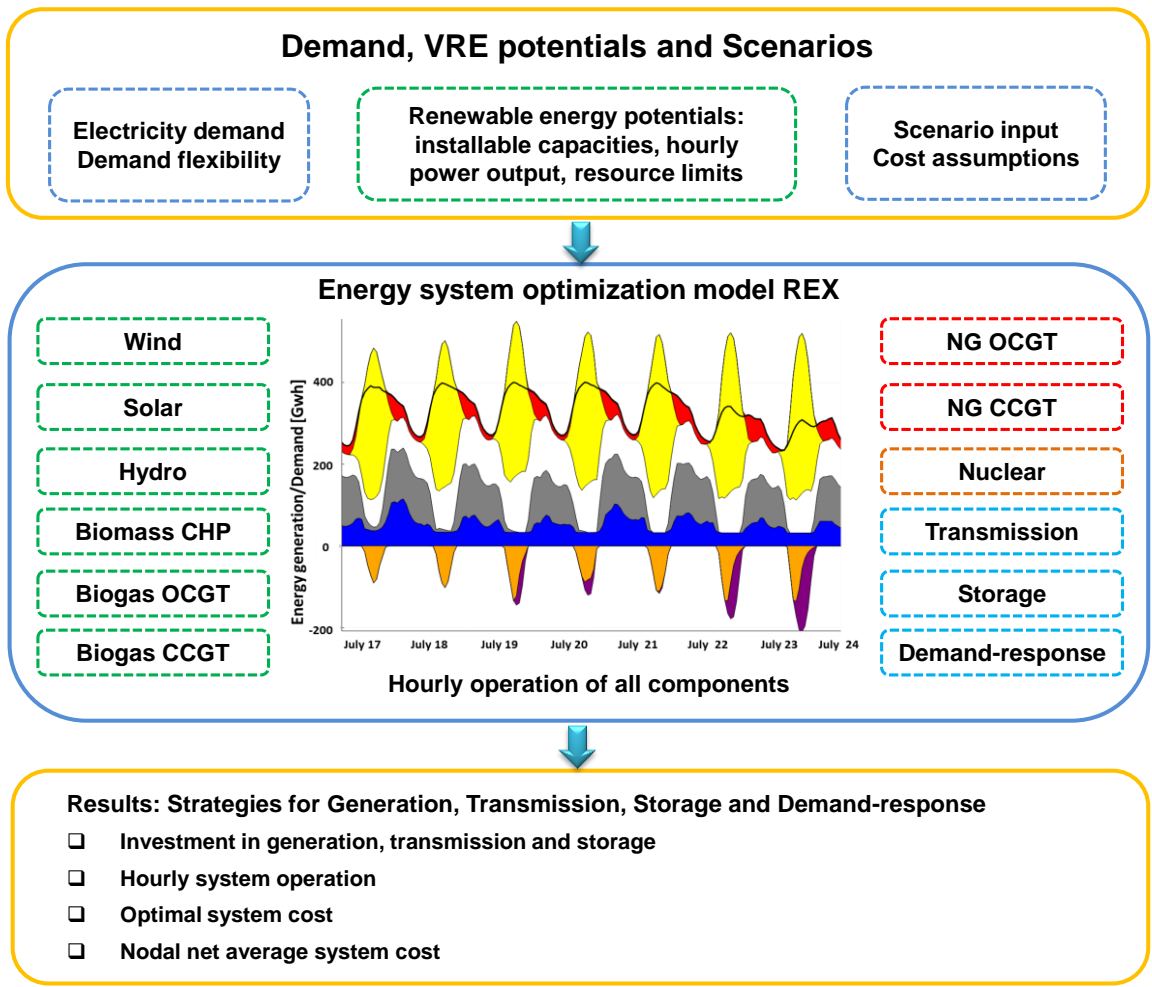

Fig. 1. Overview of methodology. CHP: Combined heat and power. OCGT: Open cycle gas turbine. CCGT: Combined cycle gas turbine. NG: Natural gas

## 2.1 Optimization model REX

The model REX is designed as a greenfield capacity expansion model that optimizes investment and dispatch of the electricity generation. Instead of looking at the transitioning pathway towards a new system, the model seeks the minimum cost portfolio for the future electricity system, a so-called overnight investment approach. The objective of the model is to minimize the total annual system cost under the constraints of meeting electricity demand, renewable energy resource potentials and a $CO_2$ emission cap. The main decision variables in the model consists of installed capacity for generation, storage, transmission, the amount of demand-response, as well as the hourly dispatch. The model REX is similar to the energy system model developed by Mattsson et al. [25] except that we have a more detailed representation for hydropower.

The nodes in the model are labeled by *r*, generation and demand-response at the node are labeled by *n*, and hours of the year are labeled by *t*. The total annual system cost incorporates annualized investment cost for thermal generation capacity $x_{nr}$, variable renewable generation capacity $y_{nr}$, transmission capacity $z_{rr'}$, storage $u_r$, and variable cost for thermal generation $\alpha_{nrt}$ and demand-response $v_{nrt}$. For VRE, transmission and storage, the variable cost is assumed to be zero. Therefore, the objective function of this linear optimization problem is formulated as

$$Min \sum_{t \in T} \sum_{r \in R} \sum_{n \in X} (C_n x_{nr} + R_n \alpha_{nrt}) + \sum_{r \in R} \sum_{n \in Y} (C_n y_{nr}) + \sum_{r \in R} (C_s u_r) + \sum_{r \in R} \sum_{r' \in R} (0.5 C_{rr'} z_{rr'})$$
$$+ \sum_{t \in T} \sum_{r \in R} \sum_{n \in M} (R_n v_{nrt}), \quad (1)$$

where $C_n$ is the annualized investment cost for generation technology *n*, $C_s$ is the annualized investment cost for storage, $C_{rr'}$ is the annualized investment cost for transmission line *rr'*, and $R_n$ is the variable cost for generation technology and demand-response. Since $z_{rr'}$ and $z_{r'r}$ represent the capacity for the same transmission line $rr'$, a coefficient of 0.5 is added to the transmission cost formula to avoid double accounting of the cost.

One main constraint for the optimization is that electricity demand has to be satisfied through generation, trade, storage, and demand-response to guarantee the security of electricity supply.

$$\sum_{n \in X} \alpha_{nrt} + \sum_{n \in Y} \beta_{nrt} + \sum_{n \in M} v_{nrt} + \sum_{r' \in R} (\eta_\gamma \gamma_{r'rt} - \gamma_{rr't}) + \sum_{n \in S} (\eta_s \epsilon_{rt} - \delta_{rt}) + h_{rt} \geq D_{rt} \leftrightarrow \lambda_{rt}, \quad (2)$$

where $\alpha_{nrt}$ is the generation of thermal power plants, $\beta_{nrt}$ is the generation of VRE, $\gamma_{rr't}$ is the electricity trade from node *r* to node *r'*, $\eta_\gamma$ is the efficiency of transmission, $\epsilon_{rt}$ is the discharge from storage, $\delta_{rt}$ is the charge into storage, $\eta_s$ is the round-trip efficiency of storage, $h_{rt}$ is the generation of hydropower, $D_{rt}$ is the hourly electricity demand, $\lambda_{rt}$ is the Karush-Kuhn-Tucker (KKT) multiplier associated with this constraint. The KKT multiplier indicates the marginal price of supplying additional demand at node *r* in hour *t* [26].

All the other constraints for the optimization problem and the details of the model are listed in Appendix A. The model was implemented in Julia using JuMP [27] and was optimized using the Gurobi [28] solver. The calculation time is between 12 and 32 minutes depending on the specific scenarios. Dell Precision 5820 Tower with Intel® Core™ i9-9900X CPU @3.50 GHz, RAM 64 GB and Windows 10, 64-bit system, is used for the implementation of the model.

**2.2 Nodal net average system cost**

The total system cost of, for instance, the European electricity system is well defined and can readily be used as indicators to compare different scenarios [6]. However, evaluating the cost of an individual country in the continental electricity system is more difficult. The generation capacity invested in a specific country may end up supplying electricity to neighboring countries. If only the capital and operational costs in each country are assessed, the importing country may be perceived as having a very low-cost electricity system. However, this is not necessarily true as the cost of the imported electricity is ignored. This problem can be avoided if countries are studied in isolation, but this obviously fails to capture the interplay with surrounding regions.

Tranberg et al. [29] introduced a method to assign the shares of capital and operational costs associated with imported electricity from generation capacities abroad to the importing countries through tracing the power flow. Pattupara and Kannan [15] incorporated revenue from electricity trade in the system cost and evaluated the effect of electricity trade on the national electricity system cost. We use a similar approach in this paper by introducing the concept of nodal net average system cost (NNASC) to represent the net electricity system cost for each node in the model. This concept incorporates the system-wide capital and operational costs of generation and transmission, profit of trade (revenue from exporting electricity less the cost of importing electricity), and congestion rent[2].

Node $r$ imports and exports electricity at the nodal price $\lambda_{rt}$, and receives the congestion rent resulting from electricity export. The capital cost of transmission infrastructures assigned to node $r$ is assumed to be proportional to the share of annual congestion rent it receives. Therefore, the annual nodal net system cost (ANNSC) for node (country or region) $r$ is

$$C_r = \sum_{t \in T} \sum_{n \in X}(C_n x_{nr} + R_n \alpha_{nrt}) + \sum_{n \in Y}(C_n y_{nr}) + C_s u_r + \sum_{t \in T} \sum_{n \in M}(C_n v_{nrt}) + \sum_{t \in T} \sum_{r' \in R}(\gamma_{r'rt} \lambda_{rt})$$
$$- \sum_{t \in T} \sum_{r' \in R} (\gamma_{rr't} \lambda_{rt}) - \sum_{t \in T} \sum_{r' \in R} (\gamma_{rr't}(\lambda_{r't} - \lambda_{rt}))$$
$$+ \sum_{r' \in R}(C_{rr'} z_{rr'} \frac{\sum_{t \in T} \gamma_{rr't}(\lambda_{r't} - \lambda_{rt})}{\sum_{t \in T} \gamma_{rr't}(\lambda_{r't} - \lambda_{rt}) + \sum_{t \in T} \gamma_{r'rt}(\lambda_{rt} - \lambda_{r't})}), \quad (3)$$

where $\sum_{t \in T} \sum_{n \in X}(C_n x_{nr} + R_n \alpha_{nrt}) + \sum_{n \in Y}(C_n y_{nr}) + C_s u_r + \sum_{t \in T} \sum_{n \in M}(C_n v_{nrt})$ is the overall generation costs in $r$ ; $\sum_{t \in T} \sum_{r' \in R}(\gamma_{r'rt} \lambda_{rt})$ is the cost of electricity import; $\sum_{t \in T} \sum_{r' \in R}(\gamma_{rr't} \lambda_{rt})$ is the revenue from electricity export; $\sum_{t \in T} \sum_{r' \in R}(\gamma_{rr't}(\lambda_{r't} - \lambda_{rt}))$ is the sum of congestion rent that $r$ receives; $C_{rr'} z_{rr'}$ is the investment cost of transmission line $rr'$ ;

---

[2] *Congestion rent* is defined as the price difference times the power flow over a transmission network constraint.

$\frac{\sum_{t \in T} \gamma_{rr't}(\lambda_{r't} - \lambda_{rt})}{\sum_{t \in T} \gamma_{rr't}(\lambda_{r't} - \lambda_{rt}) + \sum_{t \in T} \gamma_{r'rt}(\lambda_{rt} - \lambda_{r't})}$ is the share of annual congestion rent, resulting from transmission line *rr'*, allocated to *r*.

Dividing the ANNSC by the total electricity consumption in node *r* yields the corresponding nodal net average system cost (NNASC, represented by $C_r^{av}$). The NNASC manages to capture the net average system cost for an individual country or region in the interconnected electricity system.

$$C_r^{av} = \left[ \sum_{t \in T} \sum_{n \in X} (C_n x_{nr} + R_n \alpha_{nrt}) + \sum_{n \in Y} (C_n y_{nr}) + C_s u_r + \sum_{t \in T} \sum_{n \in M} (C_n v_{nrt}) + \sum_{t \in T} \sum_{r' \in R} (\gamma_{r'rt} \lambda_{rt}) \right.$$

$$- \sum_{t \in T} \sum_{r' \in R} (\gamma_{rr't} \lambda_{rt}) - \sum_{t \in T} \sum_{r' \in R} (\gamma_{rr't}(\lambda_{r't} - \lambda_{rt}))$$

$$\left. + \sum_{r' \in R} (C_{rr'} z_{rr'} \frac{\sum_{t \in T} \gamma_{rr't}(\lambda_{r't} - \lambda_{rt})}{\sum_{t \in T} \gamma_{rr't}(\lambda_{r't} - \lambda_{rt}) + \sum_{t \in T} \gamma_{r'rt}(\lambda_{rt} - \lambda_{r't})}) \right] / \sum_{t \in T} D_{rt.} \quad (4)$$

**2.3 Regions and data**

The countries included in the model are EU-28 (excluding Cyprus and Malta) plus Switzerland, Norway, Serbia, Bosnia and Herzegovina, North Macedonia, and Montenegro. The network topology of this model is shown in Fig. 2, where Sweden is divided into 4 regions, Norway is divided into 5 regions, Finland is divided into 2 regions and the rest of Europe is divided into 11 regions. In total there are 22 regions in the model and these regions are interconnected with transmission grids. Since the focus of this study is Sweden, regions close to Sweden are represented by one node each, and countries far away from Sweden are highly aggregated. We assume the future international transmission grids are high-voltage direct current (HVDC) connections and use the transport model to represent the electricity trade. The length of the transmission line is set by the distance between the geographical mid-points of each region [6]. The cost of installing converter station is 170 $/kW [6]. The current transmission capacity is based on the upper value for net transfer capacities in [30]. All modeled regions are treated as "copper plates." Thus, there is no transmission constraint within each region. The hourly electricity demand in 2014 is used as the load profile for each country and the data is obtained from ENTSO-E [31]. Demand data for the four regions of Sweden and five regions of Norway are taken from Nord Pool [32]. The demand data of Finland is divided into two regions in accordance with the share of annual electricity consumption in each region [33]. To account for potential electrification of other sectors by 2045, the electricity demand is linearly scaled up by 33% based on the reference scenarios for EU countries [34, 35] for sensitivity analysis.

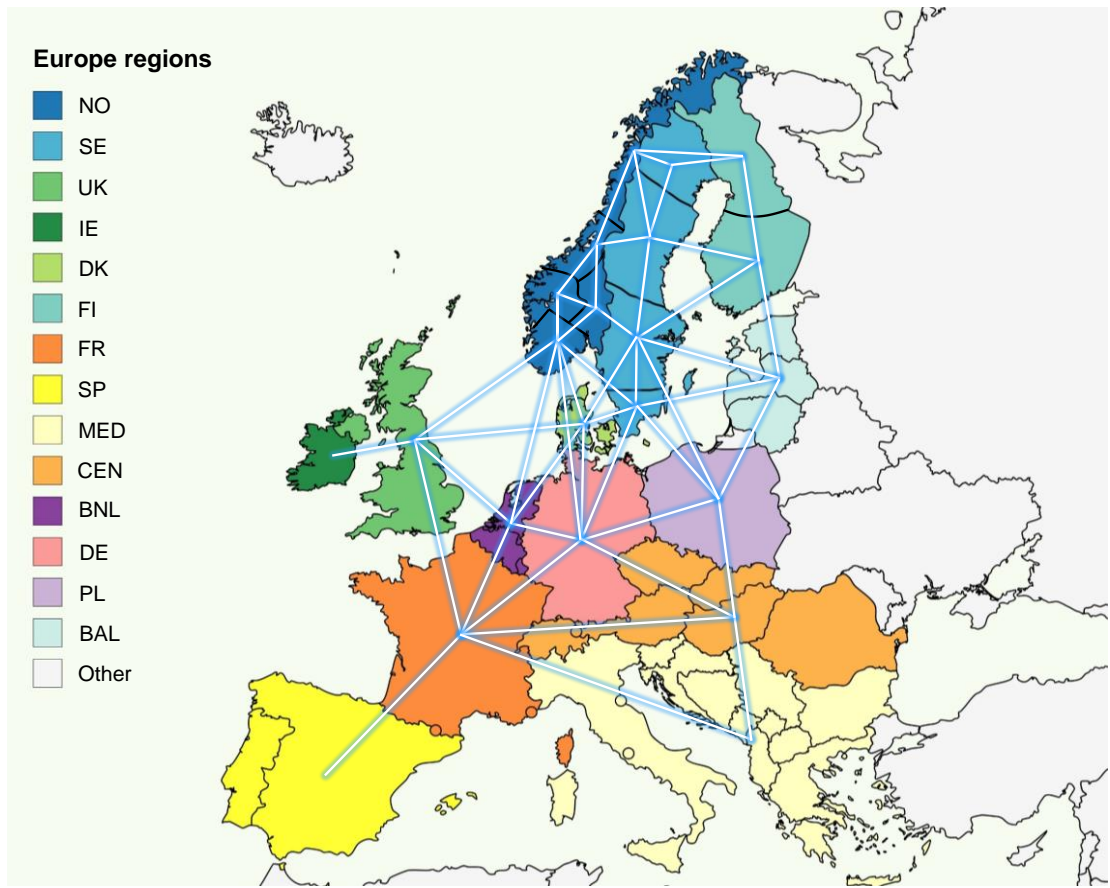

**Fig. 2**. Regions and transmission network used in the case study

In this study, new nuclear power can be invested in countries with existing nuclear fleet and the upper limit is the current capacity. Three exceptions are Germany, Switzerland and Belgium, where there are clear policies to phase out nuclear power plants [36, 37]. This means that the maximum potential nuclear energy supply is equivalent to 20% of current European electricity demand if a capacity factor of 80% is applied to nuclear power.

Due to the scarcity of the biomass primary resource, the fuel supply for biogas power plants is limited to at most 5% of the annual electricity consumption, which is approximately equal to the annual amount of biogas that could be produced from manure, agricultural residues and waste. The fuel consumption and capacity of biomass CHP plants are kept at the current level, and the electricity production follows the heat demand pattern in 2014. The biomass CHP capacity is calculated based on the total CHP capacity of each country, and the value is proportional to the share of biomass in the total primary energy supply for CHP. CHP in Sweden is used for both district heating and industrial use, and the production follows the current pattern. CHP power plants in the region SP and MED (see Fig. 2) are mainly for industrial use, and the output is evenly distributed throughout the year. In the rest of Europe, with the relatively low share of biomass in CHP fuel for most countries, we assume all the biomass CHP plants are used for district heating, and the monthly production follows the pattern of CHP district heating in Sweden.

For the storage option in the model, the cost for battery is used as reference. However, the storage may be any form of storage with a similar cost structure. Liberal assumptions on storage and low costs for storage decreases the system cost for a fully renewable electricity system (without nuclear power) [38]. Therefore, in order to create a setup which is potentially favorable for nuclear power to be a cost-effective option, pumped hydro storage is not considered in this study. For the same reason, the capacities of reservoir hydropower (hydro reservoir) and run-of-river hydropower (hydro RoR) are kept at the current level. This is also due to environmental regulations, which are not likely to change dramatically within the next few decades. The capacities for hydro are taken from ENTSO-E statistics [39]. The inflow for each country is based on reference [6] and this value is divided into reservoir and RoR inflow which is commensurate with the share of installed hydropower capacity. This study uses data from 2003 for hydro inflow, with an annual value of 439 TWh for Europe. The year 2003 was a dry year in Sweden [8], so this represents a conservative assumption for the contribution from hydropower. For Sweden, the hydro storage level is 34 TWh and the capacity is 16 GW. The minimum environmental flow [40, 41] of hydro reservoir is set to 5% of the mean annual inflow to satisfy the downstream ecosystem and human needs for water.

Demand-side management is one of the variation management options in this study. Specifically, the price-responsive demand curtailment (demand-response) is adopted. In a given time period, the aggregated consumers can curtail up to 5 % of the demand at the cost of marginal value of electricity consumption, see Appendix A for more details. The potential of demand-response in the industrial sector is around 3.4% for the Scandinavian area [42]. Since a large share of the industry in Scandinavia is located in Sweden, we scale the demand-response potential to 5% for Sweden. Demand-shifting is not considered in this study.

The input data for VRE is calculated based on the GIS model of Mattsson et al. [25]. The assumptions on wind and solar photovoltaic densities (W/m$^2$) and available land are shown in Table 1. The modeled subregions are divided into pixels (0.01°x0.01°). To better capture the weather conditions and represent the corresponding capacity factors for wind and solar power, wind and solar technologies are divided into five classes based on the resource quality. Solar irradiation is used to calculate the capacity factor profiles with the assumption that the PV technology is fixed-latitude tilted and the wind speed is translated into capacity factors based on the power curve for a typical wind farm with Vestas 112 3.075 MW wind turbines. The capacity factors are calculated using solar irradiation and wind speed from the ECMWF ERA5 database [43] and Global Wind Atlas [44]. The available land is given as a percentage of the suitable land, namely the total land less the populated areas, natural parks, lakes, mountains, etc. Mattsson et al. [25] applied a population density threshold of 150 capita/km$^2$ for populated areas. By contrast, we adopt a population density threshold of 75 capita/km$^2$ in this study to represent a more conservative estimate on the potential contribution from VRE resources. All the data for VRE profiles are based on the data source in 2018.

**Table 1** Assumptions on capacity limits for wind and solar photovoltaic. The density is the power output per unit area of a typical solar or wind farm.

|  | Solar Photovoltaic | Wind Onshore | Wind Offshore |
|---|---|---|---|
| Density [W/m$^2$] | 45 | 5 | 8 |
| Available land [%] | 5% | 8% | 33% |

The $CO_2$ emission constraint is 10 g/kWh, which is equivalent to a 98% reduction in $CO_2$ emission compared with the 1990 value for the electricity sector in Europe. The emission factor for natural gas is 198 g$CO_2$/kWh heat. The cost data used to calculate the annualized costs are summarized in Table 2. The parameters are based on the projections of the cost for 2040 and they are mainly taken from [45]. The investment cost is then converted to net present value with a 5% discount rate.

**Table 2** Cost data and technical parameters

| Technology | Investment cost [$/kW] | Variable O&M costs [$/MWh] | Fixed O&M costs [$/kW/yr] | Fuel costs [$/MWh fuel] | Lifetime [years] | Efficiency/ Round-trip efficiency |
|---|---|---|---|---|---|---|
| Natural gas OCGT | 460 | 5 | 17 | 35 | 30 | 0.4 |
| Natural gas CCGT | 920 | 6 | 20 | 35 | 30 | 0.6 |
| Biogas OCGT | 460 | 5 | 17 | 70 | 30 | 0.4 |
| Biogas CCGT | 920 | 6 | 20 | 70 | 30 | 0.6 |
| Biomass CHP[a] | 3500 | 0 | 100 | 50 | 25 | 0.25 |
| Nuclear | 4700[b] | 0 | 120 | 10 | 60 | 0.4 |
| Onshore wind | 1090[c] | 0 | 40 | n/a | 25 | n/a |
| Offshore wind | 2880 | 0 | 90 | n/a | 25 | n/a |
| Solar | 690 | 0 | 30 | n/a | 25 | n/a |
| Hydro reservoir | 2300 | 0 | 25 | n/a | 80 | n/a |
| Hydro RoR | 3450 | 0 | 70 | n/a | 80 | n/a |
| Transmission | 460[d] $/MWkm | 0 | 0 | n/a | 40 | 0.95 |
| Batteries | 220[c] $/kWh | 0 | 0 | n/a | 10 | 0.9 |

a. IEA ETSAP. [46]
b. NREL. [47]
c. Sepulveda et al. [17]
d. Schlachtberger et al. [6]

## 2.4 Scenarios and sensitivity analysis

Four base scenarios are analyzed concerning the availability of nuclear power in the Swedish electricity system and the possibility of expanding interconnecting transmission grids between neighboring regions.

1) NoNUC-Fix: Nuclear power is not a technology option in Sweden, and interconnections are kept at the current level.
2) NUC-Fix: The model may invest nuclear power in Sweden, and interconnections are kept at the current level.
3) NoNUC-Exp: Nuclear power is not a technology option in Sweden, and interconnections may expand.
4) NUC-Exp: The model may invest nuclear power in Sweden, and interconnections may expand.

We conduct sensitivity analyses for different cost combinations of nuclear power, transmission, and storage to account for uncertainties of future technology costs. Three levels of costs are assigned to each of the three technologies: "Low," "Medium" and "High." There are 27 cost combinations. In addition, we analyze the impact of higher cost for wind and solar and higher electricity demand on the net system cost and optimal investment of nuclear power in Sweden. Furthermore, we evaluate the economic impact of varying flexibility in demand-response: "No demand-response," "5% demand-response" and "10% demand-response." In total, there are 144 scenarios. Detailed cost assumptions, the flexibility of demand-response and an overview of the modeled scenarios are listed in Appendix B.

The reason why nuclear power, transmission, storage, wind, solar, electricity demand and demand-response are selected for sensitivity analysis among all input parameters is threefold: First, these parameters are assumed to be significantly important for the development of a VRE based system; Second, there are rather large uncertainties attributed to these costs [48-56]; Finally, including a wide range of assumptions about cost and technology allows us to investigate the breadth of conditions under which nuclear power can play a role in the future electricity system.

## 3. Results

### 3.1 Nodal net average electricity system cost

The availability of nuclear power has little impact on the nodal net average system cost (NNASC) for Sweden in a future decarbonized European electricity system. The value of NNASC remains stable both under the assumption that there is no expansion of the transmission grids (Fix), and under the assumption of optimal transmission expansion (Exp), see Fig. 3a. Furthermore, the NNASC for Sweden increases by 2% with transmission capacity expanding from the current level (Fix) to the optimal value (Exp). As illustrated in Fig. 3b, the composition of costs and revenues in the NNASC differs chiefly between the scenarios of current transmission (Fix) and scenarios of optimal transmission expansion (Exp). With current transmission, Sweden is a net importer (in monetary terms), but when transmission is expanded optimally, Sweden is a net exporter, and the cost for generation increases as well.

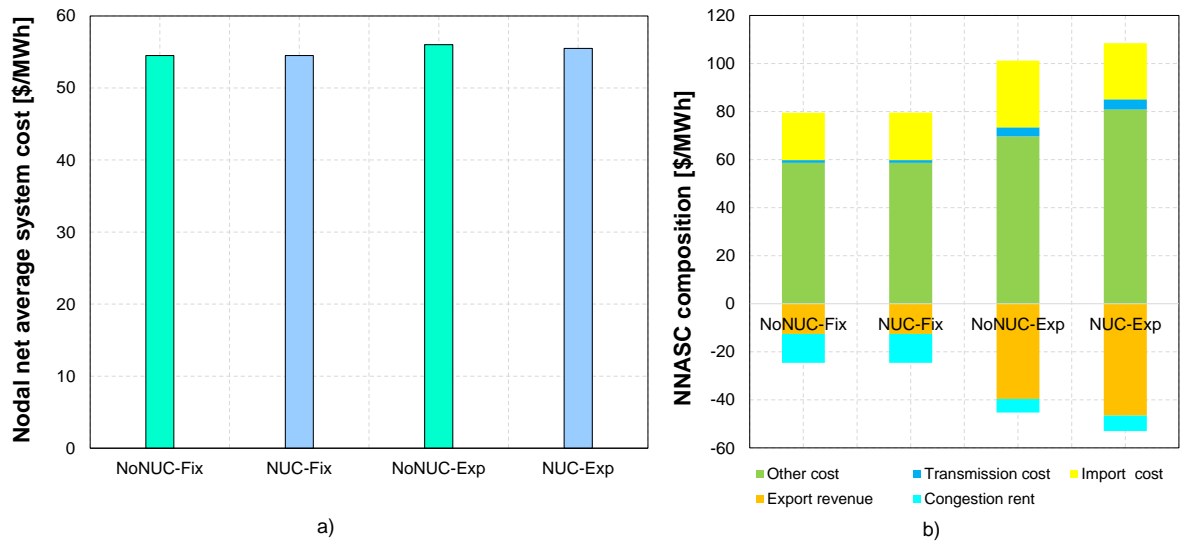

Fig. 3. Results on cost from the modeling of the base case. a) Nodal net average system cost for Sweden. b) Nodal net average system cost composition for Sweden. Since the costs of storage and demand-response are very low for Sweden, the other cost in Fig. 3b) mainly refers to generation cost.

Although Sweden is a net importer (in energy terms) in the current transmission cases (see Fig. 4b), there is still a profit, as the revenue from electricity export together with the congestion rent offset the cost of electricity imports. This shows that Sweden imports electricity when it is cheap and exports when it is expensive. The main reason for this is that Sweden has a large amount of reservoir hydropower, which enables export when renewable power supply is scarce in Europe. In the optimal transmission cases, Sweden gains even more from trade, as shown in Fig. 3b. When nuclear power is available (NUC-Exp), this effect is further enhanced, with greater net exports than in the case without nuclear power (NoNUC-Exp). The transmission cost in the 'Exp' scenarios is approximately four times as high as that in the 'Fix' scenarios. The transmission cost increase is primarily the result of the expansion of transmission capacity. In addition, with more electricity exported in the 'Exp' scenarios, Sweden is responsible for a larger share of transmission cost.

### 3.2 Capacity and energy mix for Sweden

The generation capacity- and energy mix for Sweden are shown in Fig. 4a and 4b. There is nearly no nuclear power in the capacity mix when the transmission capacity is fixed at the current level. Therefore, the capacity mix is almost the same for scenario NoNUC-Fix and NUC-Fix. The reason there is no nuclear power in the optimal capacity mix is twofold: First, with the cost assumptions in the base scenarios, the levelized cost of electricity (LCOE) of onshore wind power in Sweden is lower than that of nuclear power. Second, the abundant hydropower resources in Sweden provide variation management for wind power, as do electricity imports from Norway, thus avoiding cost escalation when there is a high penetration level of VRE in the system. If transmission is instead allowed to expand (NUC-Exp), 3.2 GW of nuclear power is installed, which is roughly equivalent to one third of the current nuclear power capacity in Sweden. With optimal transmission expansion, Sweden can provide more flexibility to the European electricity network to deal with the intermittency of renewable power supply and reduce the system-wide cost. Therefore, nuclear power is invested in scenario NUC-Exp, as the installment of nuclear power allows Sweden to export more flexibility, which is in line with the trade analysis in 3.1.

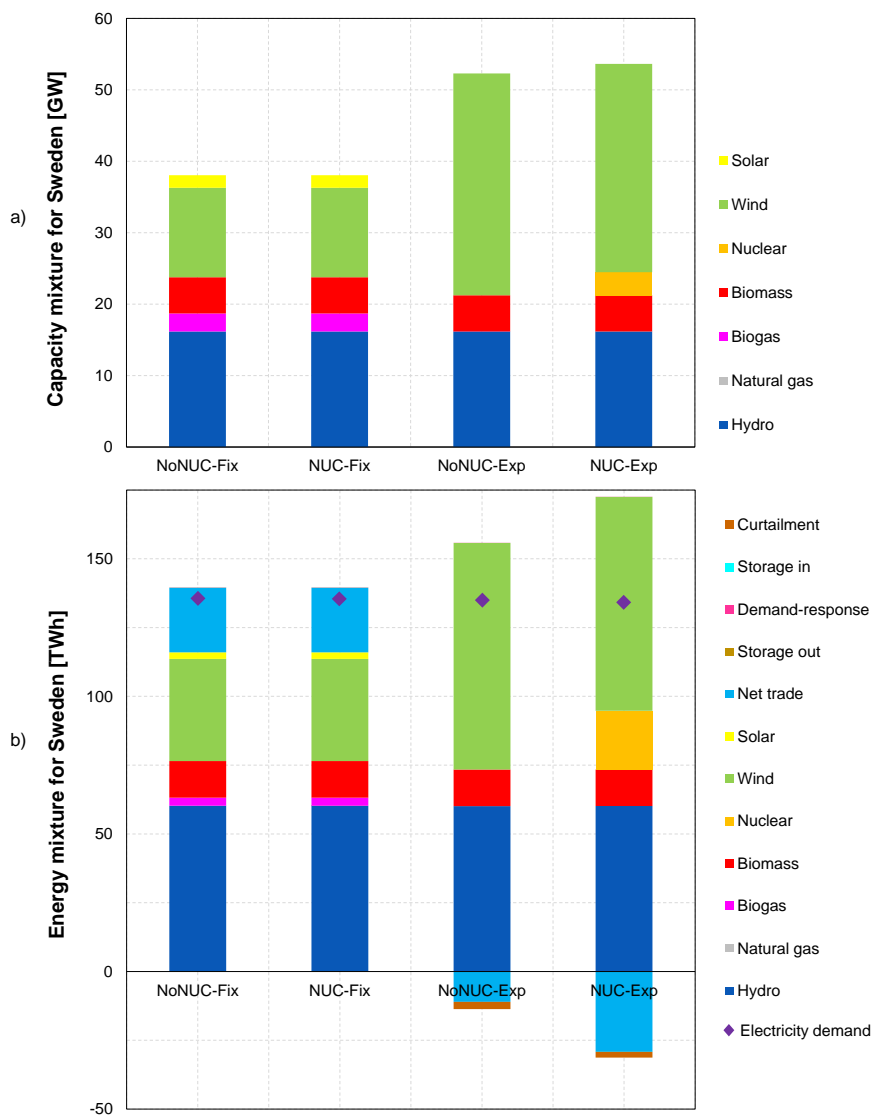

Fig. 4. Results on generation capacity and energy from the modeling of the base case. a) Capacity mixture for Sweden. b) Energy mixture for Sweden. There are nearly no investments in natural gas power plants, storage or demand-response.

Hydropower- and biomass CHP capacities are assumed to be constant due to environmental regulations and heat demand, respectively. Across all the scenarios, the generation capacity is dominated by wind (mainly onshore wind) and hydropower, see Fig. 4a. This is mainly because, in Sweden, onshore wind power is the cheapest technology to invest, and the flexibility provided by abundant hydropower can facilitate the integration of wind power. Notably, the wind power capacity is around 30 GW in the optimal transmission cases (Exp), more than twice as much as that in the current transmission cases (Fix). The wind power capacity increases with the expansion of transmission grids, as wind power in Sweden can contribute to smoothing the variation of large-scale VRE in Europe through increased electricity trade.

With current transmission connections (Cases -Fix), 1.8 GW of solar and 2.5 GW of biogas power plants are invested. By contrast, the installed capacity for solar and biogas is almost zero in the optimal transmission cases (EXP). The main reason is that when transmission is restricted, electricity imports are limited. The system has to invest in solar and biogas power to complement wind, hydropower, and CHP to satisfy the power demand locally. With optimal transmission extensions (Cases -Exp), the system can instead rely on the combination of more cost-effective wind power and trade of variations through extended transmission grids. Due to the stringent $CO_2$ cap in Europe, there is very little room for the use of natural gas. Still, due to the good reservoir hydro resources in Sweden, nearly no storage is invested.

Similar to the capacity mix, the energy mix for Sweden is dominated by wind and hydro across all base scenarios, see Fig. 4b. Nuclear energy accounts for 15% of the annual electricity generation in the NUC-Exp scenario, while this value is virtually zero in all other scenarios. In consistency with the tendency of wind power capacity in the optimal capacity mix, the increase of wind energy goes hand in hand with the extension of transmission grids. In addition, with the transmission capacity expanding from the current level (Fix) to the optimal value (Exp), Sweden shifts from a net electricity importer to a net electricity exporter. Due to high variable cost, demand-response is activated only when there are poor wind and solar conditions and all the dispatchable power plants are running at full capacity. The total energy from demand-response is less than 0.1% of the annual electricity demand in all the scenarios.

The transmission capacity for the connections to Sweden increases from 43 GW in the current transmission cases (Fix) to nearly 73 GW in the optimal transmission cases (Exp). The expansion is mainly on the transmission lines from Northern Sweden to Southern Finland, from Southern Sweden to Denmark, Germany and Poland. The average transmission capacity factor ranges from 50% to 62%. The expansion of transmission grids enables better utilization of the Nordic hydropower to supply flexibility to neighboring regions to balance the mismatch between fluctuating renewable energy generation and demand and reduce the overall system cost for Europe.

### 3.3 Sensitivity Analysis

In the base scenarios, the reduction of NNASC due to the availability of nuclear power ranges from 0% to 0.7% depending on the level of transmission capacity. We further conduct sensitivity analysis of the nuclear power cost to see how the cost of nuclear power affects its investment in the electricity system. We run 144 scenarios with different combinations of nuclear power, storage, transmission, wind, solar costs, electricity demand and demand-response to evaluate what is more cost-effective to invest in a highly renewable electricity system, nuclear power plants capable of flexible operation or variation management strategies (trade, storage and demand-response).

Figure 5 shows how nuclear power, storage and, transmission costs affect the difference in NNASC for Sweden with nuclear power relative to a system without nuclear power. As can be seen in Figure 5a, regardless of the cost-parameters, the economic benefit of nuclear power for Sweden is zero or very limited (3.2%) given the present transmission capacity. For the cases of optimal transmission, the cost difference between nuclear and non-nuclear scenarios ranges from 0% to 8%, see Figure 5b. The upper range of cost reductions are achieved, as expected, when the cost of nuclear power is low. Furthermore, the benefit of investing in nuclear power increases with higher storage cost, as more costly storage increases the cost of a highly renewable electricity system, but investments in nuclear power in Sweden enable more exports from Sweden to smooth the variations in the European electricity system and reduce the system-wide cost. In contrast, the cost of transmission has a minor effect on the potential for nuclear power to reduce system cost.

Higher electricity demand promotes the economic prospects for nuclear power investment in Sweden. If electricity demand is increased by 33% (Case-High electricity demand), the inclusion of nuclear power reduces the NNASC up to 10% for Sweden. Similarly, higher wind cost enhances the benefits of allowing nuclear power in Sweden and the corresponding cost reduction is 3.1% for fixed transmission cases and 4.1% for optimal transmission cases. On the contrary, the impact of higher solar cost is minor, with a maximum system cost reduction of 1.7% if nuclear power is included. The same holds for the amount of demand-response (see Appendix C).

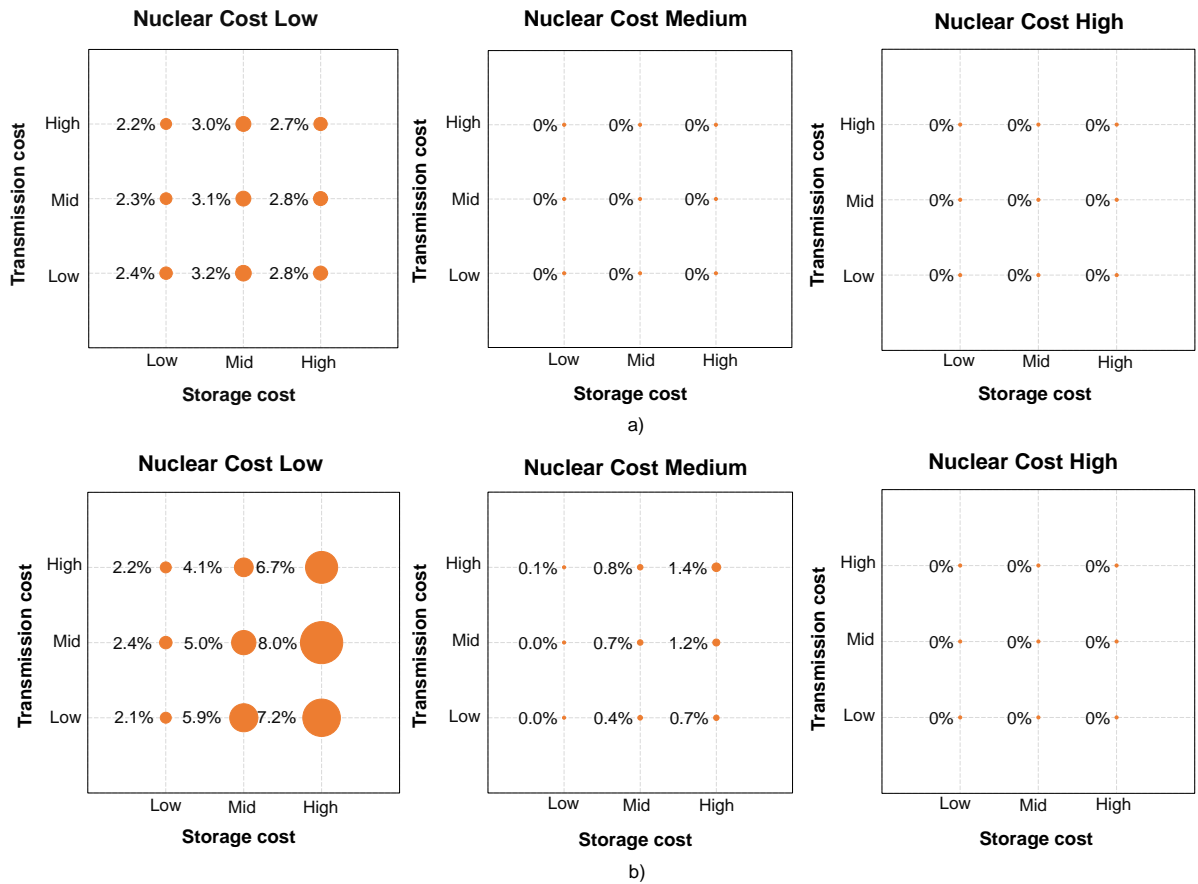

Fig. 5. The decrease of NNASC for Sweden with nuclear power compared to the case without nuclear power, using varying assumptions for nuclear power, storage and transmission costs. a) Cost difference between cases NUC-Fix and NoNUC-Fix. b) Cost difference between cases NUC-Exp and NoNUC-Exp.

## 4. Discussion

Through investigating the cost of the future low-carbon electricity system for Sweden under different scenarios, we found that the nodal net average system cost reduction due to the availability of nuclear power ranges from 0% to 8%. The upper end holds when nuclear power cost is low, transmission capacity is optimal, and storage cost is relatively high. In this case Sweden may invest in nuclear power and sell electricity to its neighboring countries at a high price, thereby making profits. With the current transmission capacity, or with moderate to high investment cost for nuclear power, the economic benefit is minor for Sweden as a country to invest in nuclear power.

Notably, there are large uncertainties in the future investment cost for nuclear power and this cost varies significantly by country (high in Europe and USA, relatively low in Asia) [48, 49, 54-58]. The cost of the two 3$^{rd}$ generation nuclear power plants (Olkiluoto 3 and Flamanville 3) currently under construction in Europe is estimated as high as 10000 $/kW [59]. Considering these nuclear power plants are the first-of-a-kind projects in Europe, we use a lower value 7000 $/kW [17] as high investment cost for this study. The medium cost, 4700 $/kW [47], represents a future cost for nuclear

power. The low cost, 3500 $/kW, is less than two thirds of the projected value for Europe today [57]. Still we use this value to represent an optimistic future where the 3$^{rd}$ generation nuclear power cost can be reduced through international standardization and the mass production of nuclear power plants.

### 4.1 Nodal net average system cost

The European electricity network is an integrated system in which the generation capacity has a diversified distribution. Countries can satisfy domestic demands through importing electricity from neighboring countries and pay for the imported electricity. The conventional system-cost concept based on generation and transmission costs cannot represent the electricity system cost for an individual country, as the effect of trade is not considered. This problem can be solved by isolating a country and not allowing for trade. However, this will entail misleading results, as electricity trade is important for power supply and variation management for renewable power systems [4-6]. Tranberg et al. [29] investigated the allocation of capital and operational costs to each node in a highly renewable power system but did not include benefits from trading electricity. Pattupara and Kannan [15] incorporated trade revenue in system cost and observed that international trade is important for the electricity system cost as trade revenue can offset the high investment costs in deep decarbonization scenarios.

To represent the system cost for an individual country in the interconnected European electricity system, we introduced the nodal net average system cost (NNASC) to incorporate trade profit and congestion rent in addition to generation and transmission costs. The composition of NNASC is shown in Fig. 3b. Note first that the transmission cost has the smallest share of the NNASC. This indicates that the mechanism adopted to allocate transmission cost does not have a large influence on the NNASC. By contrast, the revenue from electricity export, cost of electricity import, and congestion rent constitute relatively large parts of the system cost. Therefore, the allocation of electricity trade profit and congestion rent is consequential for the NNASC of individual countries. As a comparison to the method for NNASC calculation used here, we calculated the system cost for Sweden using the method proposed by Pattupara and Kannan [15] and found only a minor change of the cost estimate for Sweden due to different mechanisms applied to allocate transmission cost.

### 4.2 Comparison with studies for Sweden

Several studies have investigated the transition towards a non-nuclear electricity system for Sweden [21-23]. All these studies assessed the requirement of VRE to replace nuclear power, but only Hong et al. [21] singled out the influence of phasing-out nuclear power on the electricity system cost.

The IEA used a techno-economic cost-optimization approach to analyze a carbon-neutral scenario for the Nordic region [22]. The key finding was that nuclear power in Sweden might be replaced by 23 to 31 GW wind power, depending on the level of flexible demand from electric vehicles and heat pumps. The electricity price was estimated at around 67 $/MWh in [22]. In our study, the installed wind power capacity is around 30 GW, and the electricity price lies in the range of 65-67 $/MWh. Both the wind capacity and electricity price in our study are consistent with the results of [22]. In Söder [23], 16 GW wind power and 9 GW solar power could replace the nuclear power in Sweden. As a comparison, in our study, nuclear power may be replaced by 13 GW wind power, 1.8 GW solar power and 2.5 GW biogas power plants. The study by Söder [23] has more VRE possibly because it has a better representation (with more constraints) of hydropower and models Sweden as an isolated country.

Hong et al. [21] investigated the cost of replacing nuclear power plants with VRE in Sweden and estimated that if wind and solar power were to replace the existing nuclear power plants, the average electricity cost would be 303 $/MWh, which is significantly higher than our modeling results and the current electricity cost, 55 $/MWh [21]. The large cost difference is due to the vast expansion of wind power, with a capacity twelve times as great—154 GW—as what we find. In their study, the transmission capacity was kept at the present level, and the amount of electricity that could be imported followed historical monthly import data. Then they adopted a heuristic optimization approach to replace the decommissioned nuclear power with wind- and solar power, but the holistic power system was not optimized. In order to reveal the effect of using their method, we used the cost assumptions and $CO_2$ emission from Hong et al. as input to model REX. The test shows that the different parameter choices only explain a minor part of the difference in results. Rather, the main reason for the vast difference in results is due to the methodological choice. Instead of optimizing the whole electricity system, Hong et al. only minimized the generation capacity for wind and solar, which results in a substantial overestimate of system cost.

### 4.3 Comparison with studies for regions other than Sweden

There are a handful of studies that have assessed the cost difference of electricity system without nuclear as compared to a system with nuclear for regions other than Sweden [13, 14, 16, 17]. Jägemann et al. [13] investigated the deep decarbonization for Europe's power sector and found that the cost of decarbonization and electricity system cost together might increase by 11% if nuclear power and CCS were not included in the electricity system. Similarly, Zappa et al. [14] showed a 30% cost increase for the 100% renewable European electricity system if nuclear power and CCS were excluded. The cost difference of electricity system due to the availability of nuclear power in our study is lower compared with [13, 14]. One probable reason is that Sweden has better hydro resources than Europe on average.

Buongiorno et al. [16] focused on the role of nuclear power in the future electricity system and concluded that excluding nuclear power could increase the electricity cost significantly for deep decarbonization scenarios. They analyzed isolated regions, thus, interregional electricity trade was not included in their study. Sepulveda et al. [17] took the analysis further and studied the case with electricity trade between two US regions. The results showed that with the availability of interregional electricity trade, there is a lower cost increase even in the case of no firm low-carbon resources. In contrast, our results show that with abundant hydropower resources and full interregional electricity trade, there is no or very modest cost increase to exclude nuclear power from the Swedish electricity system.

Our modelling results for the average electricity cost of Europe range from 67 $/MWh in the case of optimal transmission up to 75 $/MWh with fixed inter-connecting transmission. These values are consistent with the results of the studies [5, 6, 18-20] on future low-carbon electricity systems in the literature. The average electricity costs of these studies are 57-103 $/MWh for Europe. Compared with these costs, the future electricity cost in Europe in our study is in the lower range.

**4.4 Limitations of the study**

Our modeling approach has three aspects that could alter the results: the spatial resolution, lack of ramping constraints on thermal technologies, and the weather data.

Regarding the spatial resolution, we have divided Sweden into four regions, Norway into five regions and Finland into two regions, while the other countries are modeled at the national level or highly aggregated. These subregions are treated as "copper plates," and the internal transmission constraints and costs are not considered. The "copper plate" assumption is likely to underestimate the cost for the system as part of the transmission cost is not accounted for. In addition, the bottlenecks of intra-country transmission grids may limit the amount of international electricity trade. Thus, if less electricity can be traded to Sweden due to internal transmission constraints in neighboring countries, more generation capacity has to be invested in Sweden, which would increase the system cost for the current transmission case. Still, Hörsch and Brown [60] observed that the effect on system cost of more detailed spatial resolution is minor for a highly renewable power system with the current transmission capacity.

Second, there are no thermal constraints, such as ramping rates for nuclear power. The ramping rate influences the speed with which nuclear power responds to the load change in the power grid [61]. With more restrictions on the ramping rate, less flexibility can be provided by nuclear power, which would primarily lead to less export revenues for Sweden if nuclear power capacity remains constant. Therefore, the lack of thermal constraints is likely to underestimate the cost for an electricity system with nuclear power. However, for a highly renewable power system, Cebulla et al. [62] found that the effect on cost from a unit commitment representation compared to a merit-order representation is minor.

Finally, we have not investigated the effect of extreme weather conditions on the future electricity system. Although we have chosen the driest year for hydro inflow in the past twenty years to represent the extremely low hydro case in Sweden, other extreme cases, such as winter nights without wind, may be more prevalent than the data we used for wind and solar [63], which would require additional back-up capacity. Therefore, we calculated the extra capacity for natural gas OCGT required to balance the system in the extreme case when there is no power production from wind and solar in Sweden, and no international electricity trade. The system without nuclear power requires 3.2 GW more natural gas OCGT than the system with nuclear power, corresponding to an increase in NNASC by 2 % for Sweden. This value constitutes the upper bound on the additional cost to ensure resilience of the electricity system, further analysis is needed for a more precise estimate. Nevertheless, guaranteeing resilience in the system does not seem to change our modeling results dramatically if nuclear power is not included in the future low-carbon electricity system for Sweden.

**5. Summary and conclusions**

In this paper, a greenfield techno-economic cost optimization model for Europe is developed to investigate the cost of a future low-carbon electricity system without nuclear for Sweden, assuming a $CO_2$ emission constraint of 10 g/kWh. With the implementation of the model, the optimal investment and dispatch of generation, transmission, storage and demand-response are achieved.

We analyze the nodal net average electricity system cost and optimal system composition for Sweden and show that:

- The nodal net average system cost for Sweden is virtually the same, irrespective of whether nuclear power is included in the electricity system or not. This implies that there is little economic rationale for Sweden as a country to invest in nuclear power if there is a transition towards a low-carbon electricity system in Europe;
- The case with best economic prospects for nuclear power investment in Sweden is when transmission capacity is optimal, combined with low cost for nuclear power and high cost for storage. In this case, the inclusion of nuclear power reduces the NNASC for Sweden by 8%. The economic rationale for nuclear power in Sweden is to enable exporting more flexibility to the highly renewable European electricity system rather than to satisfy domestic demand;
- In a highly renewable electricity system, allowing additional investment in transmission capacity would benefit Sweden through increased profits from electricity trade;

- In a future low-carbon electricity system, the nodal net average system cost for Sweden ranges from 50 $/MWh to 62 $/MWh;
- Using a heuristic approach without optimizing the whole electricity system may vastly overestimate the cost of not allowing nuclear power investment in Sweden.

We anticipate future studies with more detailed models to represent hydropower at the aggregated regional level. A better description of hydropower could validate the results of the present study. Besides, we anticipate that more case studies of other geographic areas, such as Europe, USA, and China, could confirm or reject the universality of some of the conclusions drawn from this paper.

**Acknowledgments**

The authors thank Niclas Mattsson for providing the GIS data. The authors also thank Afzal Siddiqui, Daniel Johansson and Christian Azar for helpful discussions and suggestions. This work was co-funded by the European Union's Horizon 2020 research and innovation program under the Marie Skłodowska-Curie grant agreement No: 765515 (ENSYSTRA) and Chalmers Energy Area of Advance.

**Supplementary material to "*The cost of a future low-carbon electricity system without nuclear power for Sweden*"**

Xiaoming Kan[a, *]   Fredrik Hedenus[a]   Lina Reichenberg[a, b]

a Department of Space, Earth and Environment, Chalmers University of Technology, Gothenburg, Sweden

b Department of Mathematics and Systems Analysis, Aalto University, Helsinki, Finland

* Corresponding author: kanx@chalmers.se

**Appendix A: REX model**

The mathematical representation of our model for capacity investment and dispatch of electricity generation, transmission, storage and demand-response is described below. It is a linear optimization problem with the objective of minimizing total annual electricity system cost. It employs overnight investment in a greenfield optimization approach. Instead of investigating the transition pathway towards a low-carbon electricity system, the model seeks a cost-optimal portfolio for the future electricity system under a $CO_2$ cap.

Generation technologies included in the model are wind, solar, hydropower, biomass CHP, biogas OCGT, biogas CCGT, natural gas OCGT, natural gas CCGT and nuclear power. Variation management strategies included are transmission, storage and demand-response. The full model is written in Julia using the JuMP optimization package and solved with the Gurobi solver. More details of the model, the objective function and the constraints are listed below:

**Sets**

| | |
|---|---|
| $R$ | Regions |
| $T$ | Time steps |
| $X$ | Thermal technologies |
| $Y$ | VRE technologies |
| $S$ | Storage technologies |
| $M$ | Demand-side resources |

**Parameters**

| | |
|---|---|
| $D_{rt}$ | Demand in time-step $t$ in region $r$ [$MWh$] |
| $C_n$ | Annualized investment cost for generation technology $n$ [$\frac{\$}{MW}$] |
| $C_r$ | Annualized investment cost for storage $[\frac{\$}{MWh}]$ |
| $C_{rr'}$ | Annualized investment cost for transmission line $rr'$ $[\frac{\$}{MW}]$ |
| $R_n$ | Variable cost for generation technology and demand-side resource $[\frac{\$}{MWh}]$ |
| $A_{nr}$ | Area available for VRE resource $n$ in region $r$ $[km^2]$ |
| $\rho_{nr}$ | Capacity density assigned to VRE resource $n$ in region $r$ $[\frac{MW}{km^2}]$ |
| $\eta_{n,\gamma,s}$ | Efficiency for generation technology, transmission and storage |
| $O_{nrt}$ | Output of variable quantity for VRE technology $n$ in region $r$ in time-step $t$ |
| $W_{rmax}$ | Maximum storage level of hydro reservoir in region $r$ $[MWh]$ |
| $W_{rmin}$ | Minimum storage level of hydro reservoir in region $r$ $[MWh]$ |
| $f_{irt}$ | Hydro inflow to reservoirs in region $r$ during time-step $t$ $[MWh]$ |
| $f_{irt}'$ | Hydro inflow to RoRs in region $r$ during time-step $t$ $[MWh]$ |
| $H_r'$ | Hydro RoR generation capacity in region $r$ $[MW]$ |
| $H_r$ | Hydro reservoir generation capacity in region $r$ $[MW]$ |
| $\varphi_n$ | Fraction of demand-response that belongs to segment $n$ |
| $E_n$ | Emission factor of generation technology $n$ $[\frac{gCO_2}{MWh}]$ |
| $Cap$ | Carbon cap |

**Variables**

| | |
|---|---|
| $x_{nr}$ | Capacity of thermal technology $n$ in region $r$ $[MW]$ |
| $y_{nr}$ | Capacity of VRE technology $n$ in region $r$ $[MW]$ |
| $z_{rr'}$ | Capacity of transmission between region $r$ and $r'$ $[MW]$ |
| $u_r$ | Storage in region $r$ $[MWh]$ |
| $\alpha_{nrt}$ | Generation using thermal technology $n$ in region $r$ during time-step $t$ $[MWh]$ |
| $\beta_{nrt}$ | Generation using VRE technology $n$ in region $r$ during time-step t $[MWh]$ |
| $\gamma_{rr't}$ | Exported electricity from region $r$ to $r'$ during time-step $t$ $[MWh]$ |
| $\gamma_{r'rt}$ | Imported electricity from region $r'$ to $r$ during time-step $t$ $[MWh]$ |
| $\theta_{rt}$ | Storage level in region $r$ during time-step $t$ [$MWh$] |
| $\delta_{rt}$ | Electricity into storage in region $r$ during time-step $t$ [$MWh$] |
| $\epsilon_{rt}$ | Electricity out of storage in region $r$ during time-step $t$ [$MWh$] |
| $w_{rt}$ | Hydro reservoir storage level in region $r$ during time-step $t$ [$MWh$] |
| $f_{ort}$ | Hydro outflow from reservoirs in region $r$ during time-step $t$ [$MWh$] |
| $f_{ort}'$ | Hydro outflow through the turbine in region $r$ during time-step $t$ [$MWh$] |
| $f_{ort}''$ | Hydro outflow bypassing turbine in region $r$ during time-step $t$ [$MWh$] |
| $h_{rt}$ | Hydro generation in region $r$ during time-step $t$ [$MWh$] |
| $h_{rt}'$ | Hydro RoR generation in region $r$ during time-step $t$ [$MWh$] |
| $\lambda_{rt}$ | KKT multiplier of the demand constraint in region $r$ during time-step $t$ [$\frac{\$}{MWh}$] |
| $v_{nrt}$ | Curtailed demand of segment $n$ in region $r$ during time-step $t$ [$MWh$] |
| $C_r$ | Annual net electricity system cost for country (or region) $r$ [$\$$] |
| $C_r^{av}$ | Nodal net average system cost for country (or region) $r$ [$\frac{\$}{MWh}$] |

## 1. Objective Function

Minimize total annual system cost: thermal power capacity and generation costs + VRE capacity costs + storage cost + transmission capacity cost + demand-response cost. Therefore, the objective function is formulated as

$$Min \sum_{t \in T} \sum_{r \in R} \sum_{n \in X}(C_n x_{nr} + R_n \alpha_{nrt}) + \sum_{r \in R}\sum_{n \in Y}(C_n y_{nr}) + \sum_{r \in R}\sum_{n \in S}(C_s u_r) + \sum_{r \in R}\sum_{r' \in R}(0.5 C_{rr'} z_{rr'})$$
$$+ \sum_{t \in T}\sum_{r \in R}\sum_{n \in M}(R_n v_{nrt}). \qquad (A.1)$$

## 2. Constraints

### 2.1 Demand Balance Requirement

The electricity demand has to be satisfied through generation, trade, storage and demand-response to guarantee the security of the electricity supply.

$$\sum_{n \in X}\alpha_{nrt} + \sum_{n \in Y}\beta_{nrt} + \sum_{n \in M}\varepsilon_{nrt} + \sum_{r' \in R}(\eta_\gamma \gamma_{r'rt} - \gamma_{rr't}) + \sum_{n \in S}(\eta_s \epsilon_{rt} - \delta_{rt}) + h_{rt} \geq D_{rt} \leftrightarrow \lambda_{rt}, \qquad (A.2)$$

where $\lambda_{rt}$ is the Karush-Kuhn-Tucker (KKT) multiplier associated with the constraint. The KKT multiplier indicates the marginal price of supplying additional demand at node $r$ in hour $t$ [1].



## 2.2 CO₂ Emissions Constraints

The total CO₂ emissions are constrained by the carbon cap,

$$\sum_{t \in T} \sum_{r \in R} \sum_{n \in X} (E_n \frac{\alpha_{nrt}}{\eta_n}) \leq Cap. \tag{A.3}$$

## 2.3 Thermal Electricity

The hourly thermal generation is upper-bounded by capacity multiplies time interval, which is 1 $h$,

$$0 \leq \alpha_{nrt} \leq x_{nr}. \tag{A.4}$$

## 2.4 Variable Renewable Energy

The investment in VRE capacity is constrained by area considerations and the maximum capacity density,

$$0 \leq y_{nr} \leq y_{nr}^{max} = \rho_{nr} A_{nr}; \tag{A.5}$$

The hourly VRE generation is upper-bounded by momentary weather conditions and capacity,

$$0 \leq \beta_{nrt} \leq O_{nrt} y_{nr}. \tag{A.6}$$

## 2.5 Hydro Energy

For hydro reservoir, due to environmental concern, the water in the reservoir cannot exceed the maximum storage level $W_{rmax}$. In addition, the water in the reservoir is required to be above the minimum level. Therefore, the storage level has a lower bound $W_{rmin}$,

$$W_{rmin} \leq w_{rt} \leq W_{rmax}. \tag{A.7}$$

The reservoir storage level is also affected by the inflow and outflow,

$$w_{r,t+1} = w_{rt} + f_{irt} - f_{ort}, \tag{A.8}$$

where the constraint is circular so that the storage level in the last time-step of the year equals the storage level in the first time-step of the year.

Part of the outflow $f_{ort}''$ bypasses the turbine to balance the ecosystem and human needs for water downstream. The outflow $f_{ort}$ has to satisfy the minimum environmental flow requirement $Q^{min}$ [2][3]. The flow $f_{ort}'$ through the turbine is upper-bounded by the hydro reservoir capacity $H_r$.

$$f_{ort} = f_{ort}' + f_{ort}'', \tag{A.9}$$

$$f_{ort} \geq Q^{min}, \tag{A.10}$$

$$0 \leq f_{ort}' \leq H_r. \tag{A.11}$$

For hydro RoR, the power production is constrained by hydro inflow $f_{irt}'$ and capacity $H_r'$,

$$0 \leq h_{rt}' \leq f_{irt}', \tag{A.12}$$



$$0 \leq h_{rt}' \leq H_r'. \tag{A.13}$$

Hydro generation is the sum of hydro RoR generation and hydro reservoir generation,

$$h_{rt} = h_{rt}' + f_{ort}'. \tag{A.14}$$

### 2.6 Transmission

The electricity traded through transmission line is upper-bounded by transmission capacity,

$$0 \leq \gamma_{rr't} \leq z_{rr'}. \tag{A.15}$$

### 2.7 Storage Technologies

The energy in storage is bounded by the maximum energy that can be stored,

$$0 \leq \theta_{rt} \leq u_r; \tag{A.16}$$

The energy out from storage is bounded by the storage level in the previous time-step,

$$\epsilon_{rt} \leq \theta_{r,t-1}; \tag{A.17}$$

The energy into storage is bounded by the space left in the storage,

$$\delta_{rt} \leq u_r - \theta_{rt}; \tag{A.18}$$

The level in storage is consistent with the charging and discharging in each hour,

$$\theta_{r,t+1} = \theta_{rt} + \eta_s \delta_{rt} - \epsilon_{rt}, \tag{A.19}$$

where the constraint is circular so that the storage level in the last time-step of the year equals the storage level for the first time-step of the year.

### 2.8 Demand-side Resource

The demand-side resource adopted in this implementation is demand-response or price-responsive demand curtailment. In a given time period *t*, consumers aggregated in segment *n* can curtail their demand with the marginal value of electricity consumption (variable cost), $C_n$. The amount of curtailed demand is upper limited by the fraction of demand in segment *n*, $\varphi_n$, times the hourly demand $D_{rt}$,

$$v_{nrt} \leq \varphi_n D_{rt}. \tag{A.20}$$

No demand rescheduling is considered in this study.

# Appendix B: Sensitivity analysis

**Table B.1** Costs of variation management technologies for sensitivity analysis

| Technologies | Low | Medium (base case) | High |
|---|---|---|---|
| Nuclear [$/kW] | 3500 | 4700 | 7000 |
| Transmission [$/MWkm] | 230 | 460 | 690 |
| Storage [$/kWh] | 110 | 220 | 330 |

**Table B.2** Costs of wind and solar for sensitivity analysis

| Technologies | Medium | High |
|---|---|---|
| Wind [$/kW] | 1090 | 1400 |
| Solar [$/kW] | 690 | 900 |

**Table B.3** Scenarios of demand-response

| Scenarios | No demand-response | 5% demand-response (base case) | 10% demand-response |
|---|---|---|---|
| Number of segments | 0 | 5 | 5 |
| Variable cost[a] | 0 | 60-70-80-90-100 | 60-70-80-90-100 |
| Size of each segment [b] | 0 | 1 | 2 |
| Total price-responsive demand[c] | 0 | 5 | 10 |

a. Variable cost of demand-response in each demand segment as a percentage of the value of lost load ($1000/MWh)
b. Size of each demand segment as a percentage of hourly demand
c. Total price-responsive demand as a percentage of hourly demand

**Table B.4** Overview of the modeled scenarios

|  | Scenario | Nuclear included for Sweden | Transmission optimized | Nuclear cost | Wind cost | Solar cost | Transmission cost | Storage cost | Electricity demand [TWh] | Demand-response |
|---|---|---|---|---|---|---|---|---|---|---|
| **Base scenarios** | NoNUC-Fix | No | No | Medium | Medium | Medium | Medium | Medium | 3156 | 5% |
|  | NUC-Fix | Yes | No | Medium | Medium | Medium | Medium | Medium | 3156 | 5% |
|  | NoNUC-Exp | No | Yes | Medium | Medium | Medium | Medium | Medium | 3156 | 5% |
|  | NUC-Exp | Yes | Yes | Medium | Medium | Medium | Medium | Medium | 3156 | 5% |
| **Sensitivity analysis** | Low nuclear cost | Yes/No | Yes/No | Low | Medium | Medium | Low/Medium/High | Low/Medium/High | 3156 | 5% |
|  | Medium nuclear cost | Yes/No | Yes/No | Medium | Medium/High | Medium/High | Low/Medium/High | Low/Medium/High | 3156 | 5% |
|  | High nuclear cost | Yes/No | Yes/No | High | Medium | Medium | Low/Medium/High | Low/Medium/High | 3156 | 5% |
|  | High wind cost | Yes/No | Yes/No | Medium | High | Medium | Medium | Medium | 3156 | 5% |
|  | High solar cost | Yes/No | Yes/No | Medium | Medium | High | Medium | Medium | 3156 | 5% |
|  | High electricity demand | Yes/No | Yes/No | Medium | Medium | Medium | Medium | Medium | 4200 | 5% |
|  | Low demand-response | Yes/No | Yes/No | Low/Medium/High | Medium | Medium | Medium | Medium | 3156 | 0% |
|  | High demand-response | Yes/No | Yes/No | Low/Medium/High | Medium | Medium | Medium | Medium | 3156 | 10% |

# Appendix C: Sensitivity analysis of demand-response

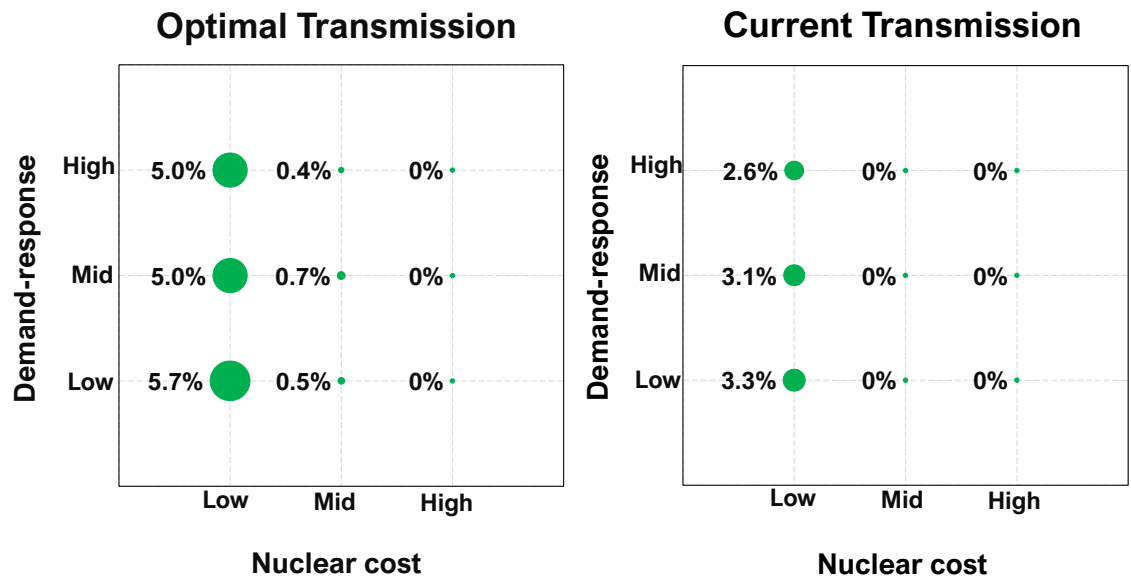

Fig. C.1. The decrease of NNASC for Sweden with nuclear power compared to the case without nuclear power, using varying assumptions for nuclear power cost and amount of demand-response.

Fig. C.1 shows the difference of NNASC for Sweden with nuclear power relative to a system without nuclear power under different combinations of nuclear power cost and demand-response. For all the scenarios, the cost difference lies in the range of 0% to 5.7%. The cost difference is more dependent on nuclear power cost than the amount of available demand-response.